\documentclass[12pt]{article}
\usepackage{graphicx}
\usepackage{hyperref}
\usepackage{amsmath}
\usepackage{scicite}
\usepackage{enumitem}

\usepackage{xr}
\newenvironment{sciabstract}{%
	\begin{quote} \bf}
	{\end{quote}}

\title{Computational multiheterodyne spectroscopy}
\author
{David Burghoff,$^{1\dagger \ast}$ Yang Yang,$^{1\dagger}$ Qing Hu$^{1}$\\
	\\
	\normalsize{$^{1}$Department of Electrical Engineering and Computer Science,} \\
	\normalsize{Research Laboratory of Electronics, Massachusetts Institute of Technology,} \\
	\normalsize{Cambridge, Massachusetts, 02139}\\
	\normalsize{$^\ast$Corresponding e-mail:  burghoff@mit.edu.} \\
	\normalsize{$^\dagger$These authors contributed equally.}
}
\date{}

\begin{document} 	
	\baselineskip24pt
	\maketitle 
	
	\begin{sciabstract}
		Dual comb spectroscopy allows for high-resolution spectra to be measured over broad bandwidths, but an essential requirement for coherent integration is the availability of a phase reference. Usually, this means that the combs' phase and timing errors must be measured and either minimized by stabilization or removed by correction, limiting the technique's applicability. In this work, we demonstrate that it is possible to extract the phase and timing signals of a multiheterodyne spectrum completely computationally, without any extra measurements or optical elements. These techniques are viable even when the relative linewidth exceeds the repetition rate difference, and can tremendously simplify any dual comb system. By reconceptualizing frequency combs in terms of the temporal structure of their phase noise, not their frequency stability, we are able to greatly expand the scope of multiheterodyne techniques. 
	\end{sciabstract}
	
	Dual comb spectroscopy is essentially the simplest possible version of multiheterodyne spectroscopy, and so they are usually considered synonymous. A frequency comb is a broadband coherent source that requires only two frequencies to fully describe it, the offset and repetition rate \cite{Udem2002}. In dual comb spectroscopy, two frequency combs are shined onto a common detector, and the heterodyne beating between different pairs of lines manifests at different radio frequencies \cite{Schiller2002,Diddams2007,Coddington2008,Coddington2010}. Though the idea is simple, the implementation is complicated by the fact that carrier-phase drift of the combs precludes coherent averaging \cite{Coddington2008,Schiller2002,Coddington2016}. When this drift is known its effect can be removed, but even measuring the absolute frequency of a comb line can be challenging. The most rigorous approach for measuring the carrier-envelope offset (CEO) directly is the f-2f technique, which requires that the comb be octave-spanning \cite{Holzwarth2000}. Another approach is to beat the comb with a stable continuous wave (CW) laser \cite{Roy2012,Ideguchi2014,Deschenes2010}. As long as the CW laser in question is moderately stable, this will provide a measurement of the comb's relative CEO. Yet another approach is to use a narrowband optical filter such as a fiber Bragg grating in order to select only a portion of the combs' optical spectrum, and to extract out the RF dual comb beating of different portions of the spectrum  \cite{Giaccari2008}. This approach requires no extra frequency references and allows for the extraction of the phase and timing fluctuations from multiheterodyne spectra, although it does require reasonably stable combs and narrowband optical filters.
	
	Recent years have also seen the development of quantum cascade laser based combs, which operate in the spectroscopically-interesting mid-infrared \cite{Hugi2012,Lu2015} and terahertz \cite{Burghoff2014,Rosch2014,Wienold2014} wavelengths. Dual comb spectroscopy based on these lasers \cite{Wang2014,Villares2014,Yang2016} is promising for enabling compact spectroscopic systems. However, performing phase and timing correction at these wavelengths is much more challenging. All of the previously-discussed approaches are viable for combs based on near-infrared mode-locked lasers, which have a well-defined time-domain profile and operate in a relatively technologically mature wavelengths. By contrast, long wavelengths possess their own challenges. For example, every additional reference channel requires an additional optical detector, but the highest-sensitivity high-speed detectors are often cryogenically cooled (e.g., mercury cadmium telluride in the mid-infrared and hot electron bolometers in the terahertz). In addition, the lasers themselves are often cryogenically-cooled, particularly in the terahertz, meaning that every extra CW laser and reference channel can greatly increase the size and complexity of the optical system.
	
	In order to overcome these challenges, we show here that nearly all of the information needed to coherently correct a multiheterodyne spectrum is contained within the RF spectrum itself, excluding information about the average absolute frequency of the two combs. These techniques are broadly applicable, applying to all sorts of combs, and as we will show they apply even when the phase and timing errors are extremely large and when little clear time- or frequency-domain structure is present. For this demonstration, we consider the case of dual terahertz quantum cascade lasers (THz QCLs) biased to a regime of marginal stability (nearing negative differential resistance). The lasers themselves are heterogeneous QCLs \cite{Rosch2014} that lase around 2.8 THz and are dispersion-compensated \cite{Burghoff2014}, while the detectors used are hot-electron bolometers and Schottky-diode mixers. This paper is focused on coherent correction; for information on the terahertz spectroscopy performed with similar lasers refer to Ref. \cite{Yang2016}.
	\begin{figure}[htbp]
		\centering
		\includegraphics[width=\linewidth]{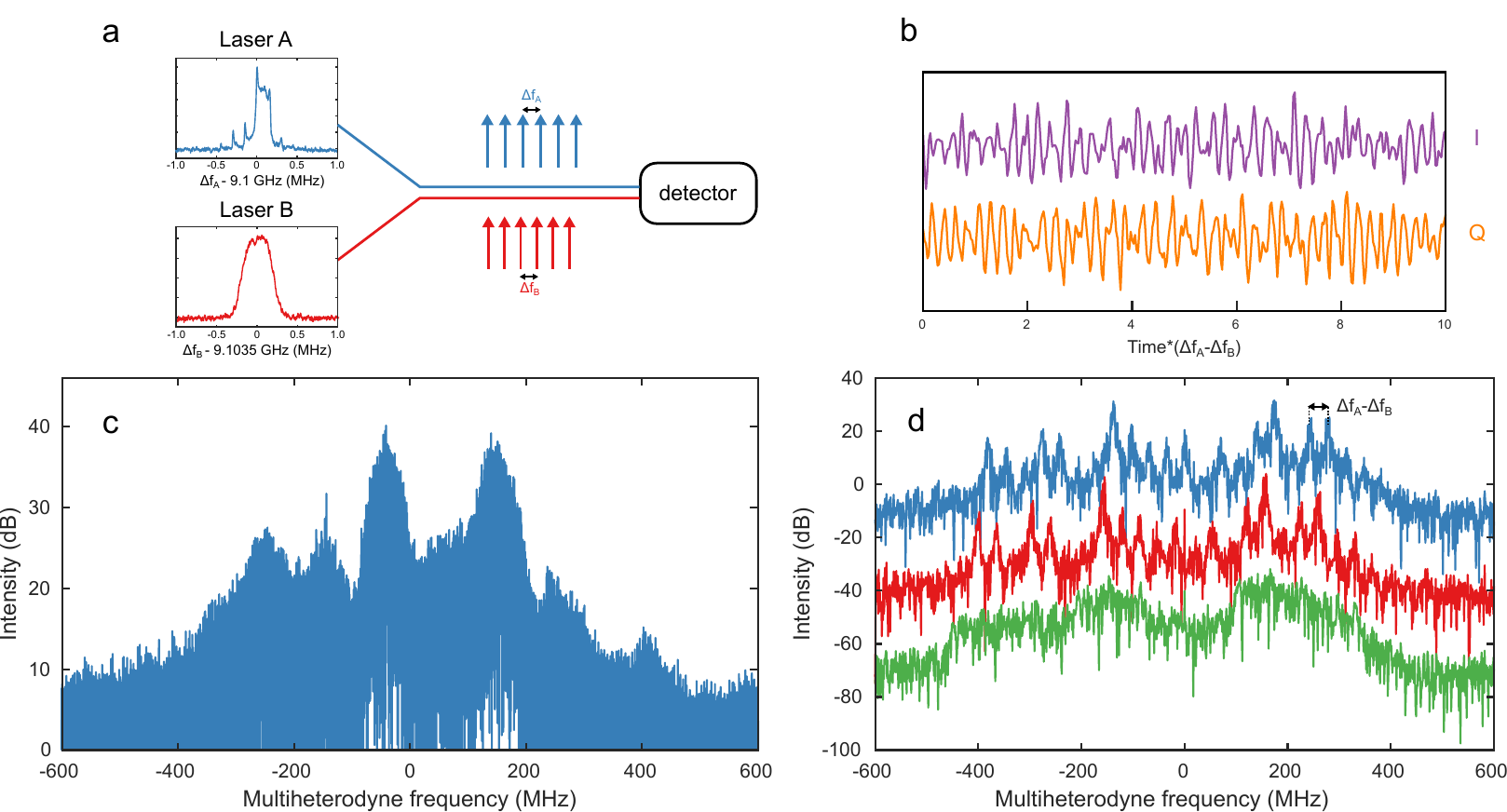}
		\caption{(a) Simplified experimental setup, showing the repetition rate beatnotes of the lasers at the biases used for this demonstration. The laser beatnotes are collected over 20 ms. (b) Quadrature components of the multiheterodyne signal in the time domain, obtained by IQ-demodulation. (c) Frequency domain multiheterodyne signal over an integration time of 100 $\mu$s. (d) Frequency domain multiheterodyne spectra over 1 $\mu$s at various times.}
		\label{fig1}
	\end{figure}Figure \ref{fig1}a shows a simplified experimental setup along with the beatnotes of the lasers, labeled A and B. These beatnotes are separated by 35 MHz, and should lead to an RF comb with repetition rate 35 MHz. However, because the lasers are operated at a bias in which they are only marginally stable, it is to some extent a stretch to call them combs. The beatnotes associated with these lasers are quite broad, and additionally, laser A possesses very clear sidebands (spaced by 140 kHz). As a consequence, the multiheterodyne signal obtained from these devices, shown in the time domain in Fig. \ref{fig1}b, is of poor quality and possesses very little evidence of the periodicity that should arise from dual combs. Consequently, in the frequency domain, shown in \ref{fig1}c, the multiheterodyne signal is broad and possesses very little evidence of a downconverted dual comb structure: over 100 $\mu$s, the duration of the recording time, all features are completely washed out. One can imagine that this is simply an issue of long term stability and that some structure might be obtained by processing the signal over shorter time intervals; spectra over 1 $\mu$s are shown in Fig. \ref{fig1}d. Though the comb structure is now evident on some spectra, it is not the case for all of them. In fact, there remain many instances in which phase instabilities completely spoil the spectrum, no matter how short the spectrum is cropped. As a result, no correction procedure that relies on interferogram cropping and alignment will succeed here \cite{Faist2016}: the signal must be corrected within the duration of an interferogram by the instantaneous phase and timing signals.
	
	Extracting the phase and timing errors from the observed multiheterodyne signal is essentially a nonlinear estimation problem. Even though we have no \textit{a priori} knowledge of these errors, we nevertheless have a model of what the RF comb should look like. Specifically, we expect it to take the form
	\begin{equation}
	y(t)=\sum_{n} A_n e^{i(\phi_0 + n \Delta \phi)},
	\label{measfcn}
	\end{equation} where $y(t)$ is the measured signal, $A_n = E^{\ast}_{n,B}E_{n,A}$ is the dual comb amplitude of the $n$th line, and $\phi_0$ and $\Delta \phi$ are the phase corresponding to the offset and repetition rate signals, $f_{0,A}-f_{0,B}=\frac{1}{2 \pi}\frac{d \phi_0}{dt}$ and $\Delta f_{A}-\Delta f_{B}=\frac{1}{2 \pi}\frac{d \Delta \phi}{dt}$, respectively. In addition, the signal itself is corrupted by additive detector noise, and the parameters are all perturbed by multiplicative amplitude noise and phase noise. Although this estimation problem may seem intractable, it has been known for some time that if the measurement was a linear function of the parameters, it would be exactly solvable by the celebrated Kalman filter \cite{Kalman1960}. In the case of a nonlinear measurement one must linearize, resulting in an inexact solution. Nevertheless, good results can still be obtained. Essentially, we are fitting the measured multiheterodyne signal to the dual comb model with the constraint that the dual comb amplitudes vary slowly. (See the Supplementary Materials for more detailed information.) With this approach we can continuously update our estimates of the offset and repetition rates without any form of cropping; this in principle makes it very amenable to real-time processing \cite{Roy2012}. Alternatively, if the data has been recorded (as in this case), it is possible to perform RTS smoothing \cite{Sarkka2008}, using future knowledge to refine the estimate and to correct for the group delay introduced by the standard filter.
	
	The physics of the comb enter primarily in the form of the multiplicative noise. Specifically, we assume that the comb complex amplitudes are perturbed only slightly at each timestep (giving them a long time constant), whereas the phase and timing errors are perturbed much more (giving them a short time constant). In other words, we assume that the comb's phase noise covariance is approximately rank-2. The Kalman filter quite naturally provides a way to test the validity of this assumption, because at every timestep it makes a prediction about what the next measurement will be. By simply comparing the measured signal to the predicted signal, we can verify the efficacy of the prediction. For all of the data shown in this paper, the prediction residual is under 8\% of the signal power.
	
	 \begin{figure}[htbp]
		  	\centering
		  	\includegraphics[width=\linewidth]{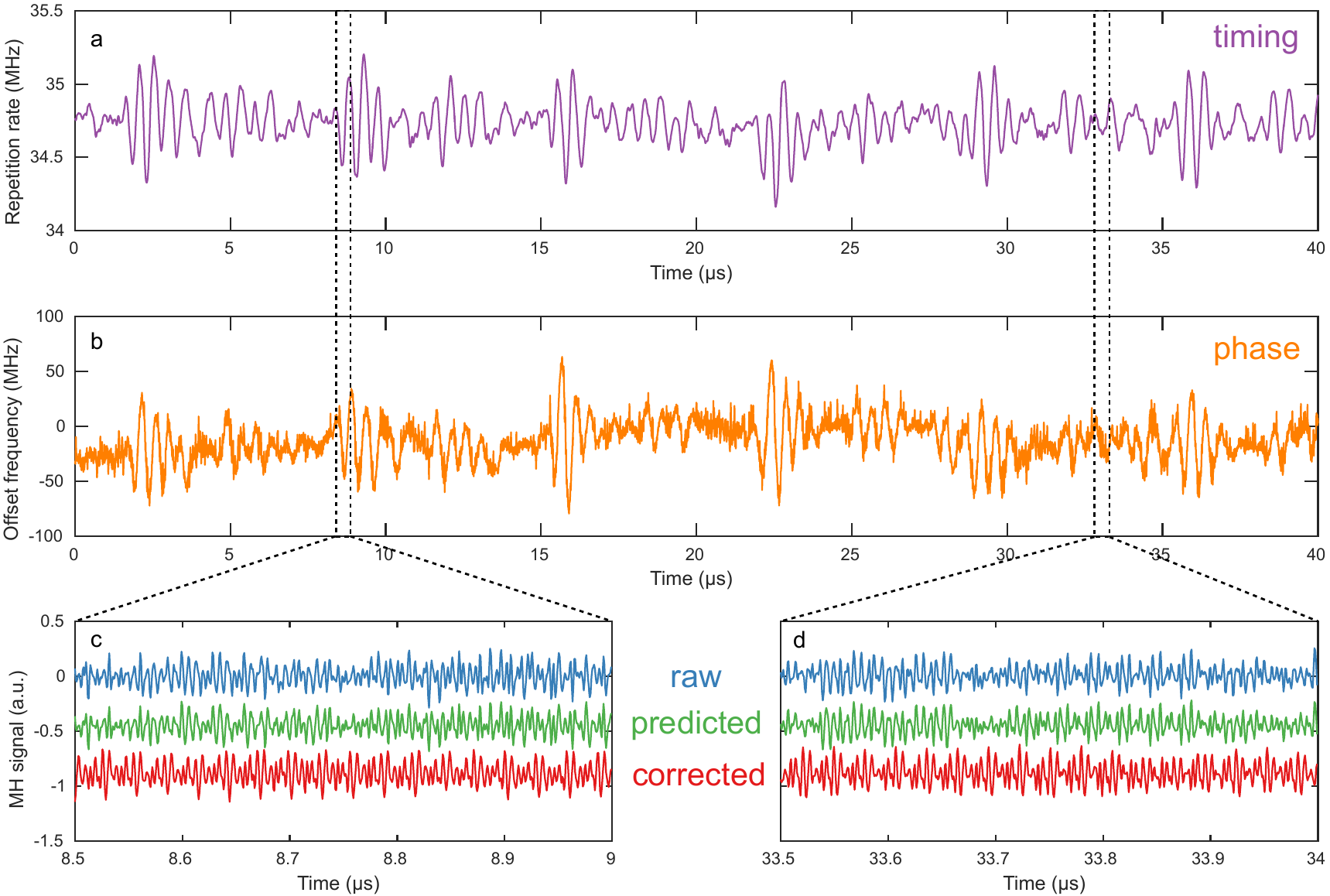}
		  	\caption{(a) Repetition rate fluctuations of the demodulated RF comb. (b) Offset fluctuations of the demodulated RF comb. (c) and (d) Raw, predicted, and corrected multiheterodyne signals during the instability and away from it.}
		  	\label{fig2}
	 \end{figure}
	
	Figure \ref{fig2}a and \ref{fig2}b show the instantaneous repetition rate and offset frequency of the RF comb in Figure \ref{fig1}. Several features are immediately apparent. The first is that both frequencies suffer a perturbation that reoccurs every 7 $\mu$s, which corresponds to the aforementioned 140 kHz sidebands evident in the beatnote of laser A. In other words, the beatnote undergoes a periodic instability that is imprinted onto the multiheterodyne spectrum. Secondly, the magnitude of the offset fluctuations (phase error) greatly exceeds the magnitude of the repetition rate fluctuations (timing error). This is not unexpected, since timing fluctuations correspond only to group index whereas phase fluctuations also depend on phase index \cite{Villares2015}. Note also that the speed of the offset fluctuations---as much as 140 MHz per 220 ns during the perturbations, roughly $0.5\Delta f$ per $1/\Delta f$---would be problematic for conventional techniques, as it implies that the short-term linewidth of a comb tooth greatly exceeds the spacing between comb teeth, a situation generally considered incompatible with dual comb spectroscopy \cite{Schiller2002,Coddington2008,Coddington2016}. Stabilization by thermal tuning is out of the question since the perturbation occurs on timescales that are too short, and even techniques that rely on measuring the beating with a CW laser would require an additional fast detector. Figs. \ref{fig2}(c) and (d) show the time-domain multiheterodyne signals before and after the phase and timing correction, both during the instability and away from it. During the instability, no clear periodicity or structure is obvious in the raw data; away from it, some periodicity is evident. In both cases, the signal predicted by the filter agrees very well with the actual data. As a result, following the phase and timing correction (discussed in Materials and Methods), the periodic comb structure is recovered.
	   
	\begin{figure}[htbp]
		\centering
		\includegraphics[width=\linewidth]{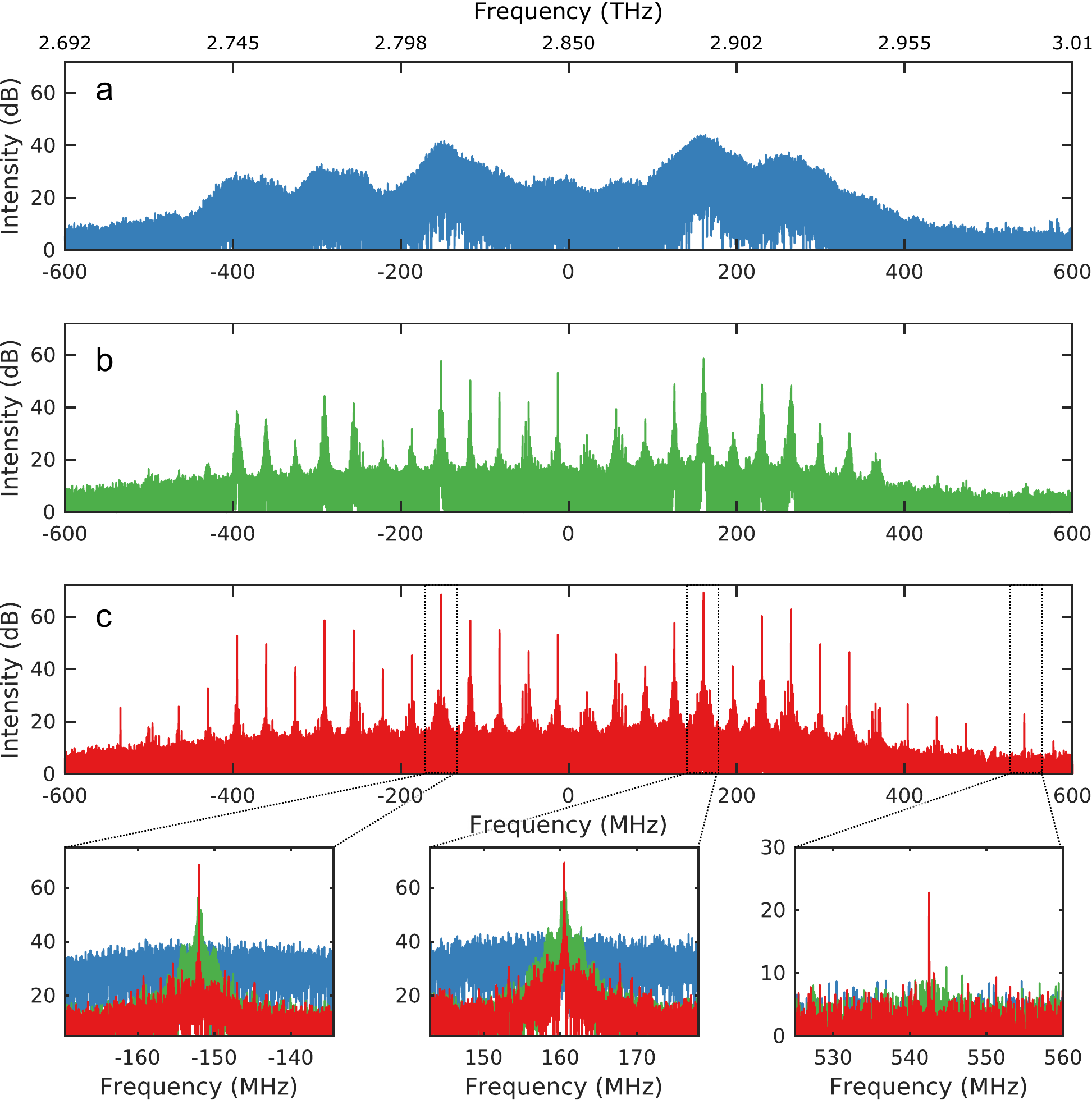}
		\caption{(a) Raw multiheterodyne data. (b) Data following phase correction. The lines in the center of the spectrum are well-localized, but the lines towards the edge are not. (c) Data following phase and timing correction. All lines are well-localized. Insets: zoomed view of three comb teeth at all three levels of correction.}
		\label{fig3}
	\end{figure}
	
	Fig. \ref{fig3} shows the results of the computational correction in the frequency domain. The raw data from before is shown in Fig. \ref{fig3}a, and once again shows no comb structure. The phase-corrected data is shown in Fig. \ref{fig3}b; because phase correction removes the average offset frequency of the signal in addition to its fluctuations, the average offset $\langle f_0 \rangle$ has been re-added to correspond with the raw data. Phase correction reveals the individual multiheterodyne comb lines, although lines near the center of the comb are better-corrected than lines near the edge because of timing fluctuations still present. Finally, Fig. \ref{fig3}c shows the phase- and timing-corrected spectra, with insets showing zoomed views of several lines. All of the lines in the spectrum have been corrected, with full-width half maxima near the uncertainty limit of 10 kHz. In fact, following the correction some lines have appeared out of the noise floor that were not apparent in the raw data, such as the one shown in the rightmost inset. By filtering the data one can verify that this is real signal (i.e., not a computational artifact), but of course given detector dynamic range limitations it may arise from detector nonlinearity rather than heterodyne beating. Even though the laser has a large disparity in mode amplitudes and large phase errors, with these techniques it is possible to perform spectroscopy \cite{Yang2016}. 
	  
	Computational phase and timing correction has advantages beyond its experimental elegance. As already shown, this approach can deal with extremely large phase-timing fluctuations. Supplementary Fig. S-1 shows an even more extreme case in which the laser is biased in an even more unstable regime, causing the comb to chaotically switch between multiple operating conditions. Even here, correction remains possible. As long as the combs are coherent in the \textit{weak} sense that the lines are evenly-spaced \cite{Burghoff2015}, with computational correction they become coherent in the \textit{strong} sense that mutually coherent dual comb spectroscopy \cite{Coddington2016} can be performed. Although we focused here on unstable combs, it is also beneficial for stable combs and even combs operated in pulsed mode \cite{Yang2016}. In addition, computational correction offers very good performance even in the case of low-signal reference measurements. For example, Supplementary Fig. S-2 shows a correction based on the multiheterodyne signal from a Schottky mixer, whose raw data has a signal-to-noise ratio (SNR) under 25 dB. Even when the mixer's noise is artificially boosted by 10 dB and little signal remains, computational correction remains informative on both the reference channel and a signal channel. One may wonder whether or not computational correction offers any multiplex advantage analogous to the one present in conventional Fourier Transform Spectroscopy \cite{Hirschfeld1976,Mandon2009}. For additive white noise, there is no advantage: the SNR of the extracted correction signal is identical between a reference comb and a reference CW laser of the same power. Even so, computational correction confers an advantage in the presence of excess phase noise, i.e. phase noise that is not offset- or timing-related. Because it uses all of the lines of the comb, it is less susceptible to the excess noise of a particular line. Indeed, we have observed that it is sometimes the case that even when most of the multiheterodyne spectrum is well-corrected, the weaker comb teeth can possess sidebands (see Supplementary Fig. S-3). These sidebands result from multiple lines sharing a mode, similar to what has been observed in microresonator combs \cite{Herr2012}, and manifest in QCL combs as a lack of coherence \cite{Burghoff2015}. If one were to use one of these lines to perform the correction instead of a clean comb line, the correction would be poor; using all of the comb's lines averages out these effects.
	
	Of course, some caveats apply. For one, because the filter needs to track the complex amplitudes of each line, this scheme is easier to apply to combs where the number of lines is not too large. This is easy to accomplish for QCL combs or microresonator combs \cite{DelHaye2007} since the number of lines is typically in the hundreds, but is computationally more complex for mode-locked lasers (where the number of lines approaches 100,000 \cite{Newbury2010}). Although this can be remedied by tracking just a few of the multiheterodyne lines (see Supplementary Figure S-4), it comes at the expense of accuracy. Additionally, because the linearized Kalman filter is sub-optimal, the correction is not perfect and always leaves a noise pedestal. For the data shown in Fig. \ref{fig3}c, the pedestal is approximately 40 dB down from the peak of each line. Lastly, because the combs are not fully-referenced, there is still of course an ambiguity that remains in the average absolute frequency of the combs \cite{Coddington2016}. This is often unimportant for spectroscopy, since the free-running linewidth of most combs is narrower than typical spectroscopic features, and the absolute frequency can be localized using an etalon or an interferometer if need be.
	
	In conclusion, we have shown that coherent multiheterodyne spectroscopy can effectively be performed using a computationally-enhanced method. Nearly all of the relevant phase and timing information traditionally acquired by separate reference channels is in fact buried within the dual comb spectrum, and can be used to self-correct or to cross-correct a spectroscopic signal channel. These approaches are particularly useful for dual comb systems based on semiconductor lasers or on microresonators, on account of their large free spectral range. Although we have used these techniques to correct comb spectra, they are very general and can apply to any coherent spectrum where a robust model of the phase-locking exists. The light source need not be comb-like, merely deterministic.

	\section*{Acknowledgments}
	The work was supported by the DARPA SCOUT program through grant number W31P4Q-16-1-0001 from AMRDEC and NSF. The authors declare no competing interests.

\bibliographystyle{Science}
\bibliography{refs}

\end{document}